\documentclass[preprint,showpacs,preprintnumbers,amsmath,amssymb]{revtex4}

\usepackage{graphicx}
\usepackage{dcolumn}
\usepackage{bm}
\usepackage{slashed}
\newcommand{\bc}{\begin{center}}
\newcommand{\ec}{\end{center}}
\newcommand{\be}{\begin{equation}}
\newcommand{\ee}{\end{equation}}
\newcommand{\ba}{\begin{eqnarray}}
\newcommand{\ea}{\end{eqnarray}}
\newcommand{\bea}{\begin{eqnarray*}}
\newcommand{\eea}{\end{eqnarray*}}

\begin{document}

\title{Sizing the double pole resonant enhancement in $e^+e^- \to \pi^0 \pi^0 \gamma$ cross section and $\tau^- \to \pi^-\pi^0 \nu_\tau\gamma$ decay}

\author{Leonardo Esparza-Arellano}
\affiliation{Instituto de F\'{\i}sica,  Universidad Nacional Aut\'onoma de M\'exico, AP 20-364,  M\'exico D.F. 01000, M\'exico}%
\author{Antonio Rojas}
\affiliation{Instituto de F\'{\i}sica,  Universidad Nacional Aut\'onoma de M\'exico, AP 20-364,  M\'exico D.F. 01000, M\'exico}%
\author{Genaro Toledo}
\affiliation{Instituto de F\'{\i}sica,  Universidad Nacional Aut\'onoma de M\'exico, AP 20-364,  M\'exico D.F. 01000, M\'exico}%

\date{\today}

\begin{abstract}
The enhancement mechanism due to the resonant properties of the $\rho$ and $\omega$ mesons, which are close in mass, are analysed when such resonances carry different momenta. Considerations from the particular process where they appear to the individual resonant features are at play for the appearance of the global resonant manifestation. In this work, we first consider the $e^+e^- \to \pi^0 \pi^0 \gamma$ process. We use the differential cross section at a given angle of emission of one of the pions, to tune the individual features of the two resonances and exhibit how both resonances combine to produce the enhancement. Then, we incorporate the $\rho^\prime$ using the information obtained from the $ e^+e^- \to \pi^0 \pi^0 \gamma$ total scattering process and show that, it becomes important thanks to the same enhancement mechanism between the $\rho$ and the $\omega$. In a second step, we use a similar approach to describe a model dependent contribution to the $\tau^- \to \pi^-\pi^0\nu_\tau\gamma$ decay, the so-called $\omega$ channel. We show that the dipion invariant mass distribution at particular angles is sensitive to the individual resonant states. We compute the interference of this  channel with the known dominant model independent contribution, and show how a better knowledge of the $e^+e^- \to \pi^0 \pi^0 \gamma$ process can help to properly account for such model dependent effects. The implication on the isospin symmetry breaking correction to tau-based estimates of the muon magnetic dipole moment is assessed.  
\end{abstract}

\maketitle
\section{Introduction}
The closeness in mass between the $\rho$ and $\omega$ resonances plays an important role in the understanding of many low energy hadronic phenomena.  The so-called $\rho-\omega$ mixing is one of the key ingredients in the proper description of the $e^+e^- \to \pi^+\pi^-$ scattering data, which is the dominant hadronic contribution for the  standard model (SM) prediction \cite{Aoyama:2020ynm,Colangelo:2022prz, Benayoun:2012wc, Davier:2010nc, Wolfe:2010gf, Wolfe:2009ts} of the muon g-2 magnetic dipole moment (MDM) \cite{Muong-2:2023cdq}. There, both mesons carry the same transferred momentum, thus their corresponding resonant features appear split in energy only by the mass difference.
A kind of similar contribution of the $\rho$ and $\omega$ resonances can be seen in the $ e^+e^- \to \pi^0 \pi^0 \gamma$ scattering process. However, in this case the mesons carry different momenta, due to the pion emission. Thus, kinematical considerations are at play, in combination with the individual resonant features, for the appearance of the global resonant manifestation. Characterize it in terms of the parameters involved may be useful to reliable describe other processes with similar characteristics, but milder or null experimental information. The $ e^+e^- \to \pi^+ \pi^- \gamma$ process is expected to be less sensitive to these features since it also includes other radiation mechanisms.

Another scenario for the appearance of both resonances with different momentum dependence can be seen in the $\tau^- \to \pi^-\pi^0\nu_\tau\gamma$ decay, which is dominated by the model independent (MI) part for soft photon emission \cite{CENPLB,CEN,Flores-Tlalpa:2005msx,FloresBaez:2006gf,FloresTlalpa:2006gs,LopezCastro:2015cja,pablo20,Masjuan:2023qsp, Chen:2022nxm}, in agreement with Low's Theorem  \cite{Low}. There, a model dependent (MD) contribution, associated to the so-called $\omega$ channel, which has a similar hadronic structure as the one in $ e^+e^- \to \pi^0 \pi^0 \gamma$, has been observed to have a relevant effect in the dipion spectrum. Properly accounting for this channel is of relevance for the isospin symmetry breaking correction on a tau-based estimate of the muon g-2 MDM prediction in the SM. While this radiative process has not been measured, experimental prospects in Belle II \cite{Belle-II:2018jsg} might offer a first insight on it.

In $e^+e^-$ scattering, the total cross section measured by SND \cite{snd2pg00,snd2pg13,snd2pg16} and CMD2 \cite{cmd22pg} experiments, at low energies, can be described considering the intermediate state to be driven by the $\rho$ and $\omega$. Refinements require the incorporation of the $\phi$ and $\rho^\prime$ mesons \cite{Moussallam:2013una,Moussallam:2021dpk,jorge,david,avalos}. In the $\tau$ decay, the considered observable is the dipion invariant mass distribution. These observables exhibit the general behavior produced by the presence of both resonances, but not the specific role of each of them, making it difficult to identify how the particular properties  combine to get the total result. This is not a trivial fact, since the processes involve a $\rho-\omega-\pi$ vertex where at least one of the mesons is off-shell. Due to energy conservation, $\rho$ and $\omega$ do not resonate at energies which differ just by the mass gap between them but larger, to account for the energy carried out by the pion.

In this work, first we analyse the behavior of the resonances in $e^+e^- \to \pi^0 \pi^0 \gamma$ considering the differential cross section for a particular angular emission of one of the pions as an additional observable. There, we use the angle to tune the individual features of the $\rho$ and $\omega$ resonances and exhibit how they combine to produce the global enhancement. Then, we incorporate the $\rho^\prime$ using the information obtained in a previous analysis \cite{avalos} of low energy observables. We show that, although it is a subdominant contribution, it becomes important thanks to the same enhancement mechanism between the $\rho$ and the $\omega$, since the kinematical energy shift allows both $\omega$ and $\rho^\prime$ to be on-shell.
Once the angular distribution is characterized, in terms of the parameters involved at the current precision, we consider the $\omega$ channel of the $\tau^- \to \pi^-\pi^0\nu_\tau\gamma$ decay, exhibiting the analogue features to $e^+e^- \to \pi^0 \pi^0 \gamma$ angular distribution in the dipion spectrum. Then, we compute its interference with the known dominant MI contribution and determine the radiative correction function $G_{EM}(t)$. We evaluate the isospin symmetry breaking correction to tau based estimates of the muon MDM from this source. We show how a better knowledge of the $e^+e^- \to \pi^0 \pi^0 \gamma$ can help to properly account for such MD contribution. At the end we  discuss the results and present our conclusions.

\section{Energy role in the form factor}
\begin{figure}[htb]
\begin{center}
\includegraphics[scale=0.6]{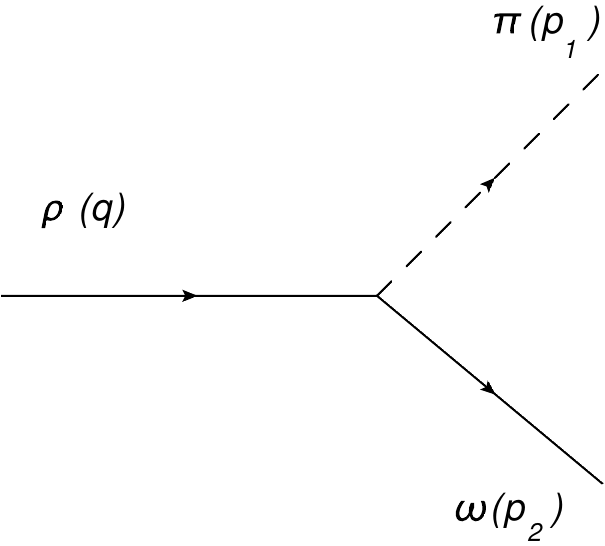}
\end{center}
\caption{$\rho\,\omega\,\pi $ interaction.}
\label{fig:rop}
\end{figure}

We can define the individual form factor associated to a vector meson ($V$) as a Breit-Wigner distribution:
\begin{equation}
    f_V[s]\equiv \frac{m_V^2}{m_V^2-s + im_V \Gamma_V},
\end{equation}
where $s$ is the kinematical variable associated and $m_V$ and $\Gamma_V$ ($V=\rho, \ \omega$) are their corresponding mass and decay width. The pole mass is not the same by definition but we will refer to them indistinctly. The particular numerical values are taken from \cite{pdg}.  Thus, for different energy dependence, we will have $f_\rho[s]$ and $f_\omega[s_1]$ for the $\rho$ and $\omega$ respectively. For the broad decay width of the $\rho$, we consider the energy dependent form $\Gamma_\rho(s)=\Gamma_\rho (m_\rho^5/s^{5/2})\,\lambda(s,m_\pi^2,m_\pi^2)^{3/2}/\lambda(m_\rho^2,m_\pi^2,m_\pi^2)^{3/2}$, where $\lambda(x,y,z)$ is the K\"allen function, while the narrow width of the $\omega$ is taken as a constant.\\
Let us consider the hadronic interaction between the $\rho$, $\omega$ and $\pi$ as shown in Fig. \ref{fig:rop}, where both $\rho$ and $\omega$ are, in general, off-shell. Although $\rho$ and $\omega$ are close in mass, it does not necessarily imply that both resonances show up close enough to each other at a given kinematical configuration.
In order to illustrate this point, let us consider the energy of the pion ($E_\pi$) in the $\rho$ rest-frame, which links $s\equiv (p_1+p_2)^2 $ and $s_1\equiv (q-p_1)^2$ variables by $s_1=s+m_\pi^2-2\,\sqrt{s}\,E_\pi$. The difference between $s$ and $s_1$ is not trivial since it depends on $E_\pi$. If the pion carries the minimal energy, $E_\pi=m_\pi$, the energy available for the $\omega$ at $s=m_\rho^2$ is $\sqrt{s_1}=0.63$ GeV, far below its mass, even considering the decay width of the $\omega$. On the other hand, the minimal energy to have the $\omega$ on-shell is $\sqrt{s}=m_\omega+m_\pi=0.92$ GeV, which is nearly $m_\rho+ \Gamma_\rho$. Thus, the appearance of both resonances requires the $\rho$ meson energy to be at least one unit of its decay width away from its mass. Phase space effects, coming from the particular process where this vertex is involved, will produce further modifications in the observables, as we show below. 
 
\section{Resonant enhancement in $e^{+}e^{-}\rightarrow \omega\pi^0\rightarrow \pi^0\pi^0\gamma$ cross section}
In the following, we will explore the behavior associated to the form factors defined above within the $e^+e^- \to \pi^0 \pi^0 \gamma$ cross section. Then, 
 we incorporate the $\rho^\prime$ contribution, which is not negligible, thanks to the same enhancement mechanism between the $\rho$ and the $\omega$, where now both $\rho^\prime$ and $\omega$  are allowed to be on-shell.
 
\begin{figure}[htb]
\begin{center}
\includegraphics[scale=0.45]{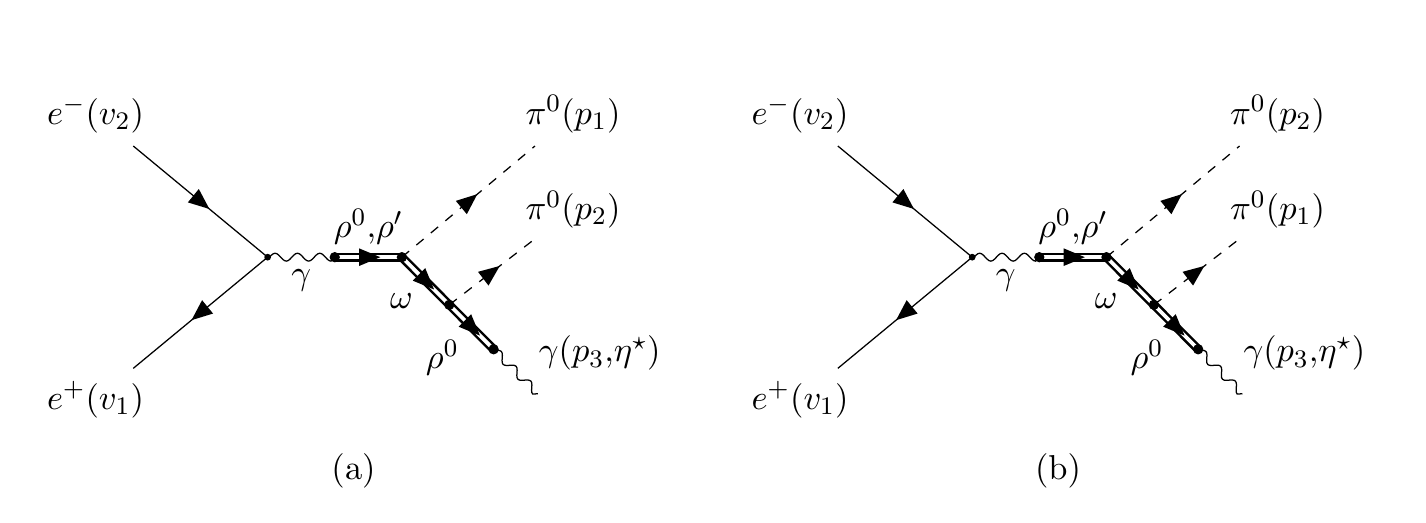}
\end{center}
\caption{The $e^+\,e^- \rightarrow \omega\pi^0 \rightarrow \pi^0\,\pi^0\,\gamma$ scattering.}
\label{fig:ppg}
\end{figure}

We follow the vector meson dominance model (VMD) to describe the coupling between the neutral vector mesons and the electromagnetic current \cite{sakurai}. The interaction among hadrons is described in an effective way, consistent with general considerations in extensions of the VMD \cite{klz,vmdbando,fujiwara,vmdx}. 
The effective Lagrangian including the light mesons $\rho$, $\pi$ and $\omega$, in addition to the $\rho^\prime$ can be set as
\begin{eqnarray}
{\cal L}&=& \sum_{V=\rho,\,\rho^\prime} g_{V\pi\pi}\,
\epsilon_{abc}\, V_\mu^a\, \pi^b \,\partial^\mu\, \pi^c 
+\sum_{V=\rho,\,\rho^\prime} 
g_{\omega V\pi}\,\delta_{ab}\,\epsilon^{\mu\nu\lambda\sigma}\,\partial_\mu\, \omega_\nu\, \partial_\lambda\, V_\sigma^a\,  \pi^b \nonumber \\
&+&
\sum_{V=\rho,\,\rho^\prime,\,\omega} \frac{e\, m_V^2}{g_V}\,V_\mu\, A^\mu.
\label{Lvmd}
\end{eqnarray}
\noindent The couplings are labeled to identify the corresponding interacting fields. In general, $V$, $A$ and $\pi$ refers to the vector meson, photon and pion fields, respectively. This approach allows to incorporate the strong interaction among hadrons and the resonances in an energy region were they can be considered as the degrees of freedom, provided the parameters can be fixed from experimental information, symmetries or low energy theorems.\\
Let us set the momenta notation (within parenthesis) for the process as:
 $e^{+}(v_{1})\,e^{-}(v_{2})\to \pi^{0}(p_1)\,\pi^{0}(p_2)\,\gamma(p_3,\eta^{*})$, where $\eta^{*}$ is the polarization vector of the photon. The process is depicted by the diagrams in Fig.~\ref{fig:ppg}, where both the $\rho$ and $\rho^\prime$ intermediate states are considered.
 Further contributions from other intermediate states, such as the $\phi$ meson or scalars, are not considered at this stage, although they may be relevant when considering this process for precision observables analysis \cite{Moussallam:2021dpk,jorge,Davier:2019can}. The amplitude for the diagram of Fig.~\ref{fig:ppg}(a) can be written as: 
\begin{equation}
 \mathcal{M}_{(a)} = \frac{e^{2}}{q^{2}}\,\Big(C_{\rho}+e^{i\theta}C_{\rho^{\prime}}\Big)\,\epsilon_{\mu\sigma\epsilon\lambda}\,q^{\sigma}\,(q-p_1)^{\epsilon}\,{\epsilon_{\alpha\beta\nu}}^\lambda\,(q-p_1)^{\alpha}\,p_3{}^{\beta}\,\eta^{*\nu}\,l^{\mu},
 \label{2pigamma}
\end{equation}
where $l^\mu\equiv-ie \bar{v}(v_1) \gamma^\mu u(v_2)$, $s=q^2=(v_1+v_2)^2$, $s_1=(q-p_1)^2$ and the global factor associated to the $\rho$ and $\rho^\prime$ intermediate states is defined in terms of the couplings and form factors by:
\begin{equation}
 C_{\rho} = \Big(\frac{g_{\omega\rho\pi}}{g_{\rho} m_\omega} \Big)^{2}\, f_\rho[s] f_\omega[s_1], 
 \hspace{0.5 cm} 
 C_{\rho^{\prime}} = \frac{g_{\omega\rho^{\prime}\pi}\,g_{\omega\rho\pi}}
 {g_{\rho}\,g_{\rho^{\prime}}m_\omega^2}\, f_{\rho^\prime}[s] f_\omega[s_1],
\end{equation}
with a relative phase $e^{i\theta}$ between both channels.
The amplitude for Fig.~\ref{fig:ppg}(b) is obtained by interchanging $p_1\leftrightarrow p_2$ momenta.

The total cross section measured by SND Coll.  \cite{snd2pg00,snd2pg13,snd2pg16} and by CMD2  Coll. \cite{cmd22pg} have been analysed following the above description, in combination with a larger set of observables, to determine the model parameters consistency region \cite{avalos}. The parameters relevant for our purposes, are listed in Table \ref{parameters1}.

\begin{table}[htb]
\begin{center}
\begin{tabular}{lc}
     \hline
     \hline
     Parameter & Value \\ \hline 
    $g_{\rho}$  &  4.962   $\pm$    0.093\\
    $g_{\omega}$  &  16.652   $\pm$    0.473\\
    $g_{\rho^\prime}$  &  12.918 $\pm$   1.191 \\ 
    $g_{\omega\rho\pi}$ (GeV$^{-1})$ & 11.314 $\pm$ 0.383\\ 
    $g_{\omega\rho^\prime\pi}$ (GeV$^{-1})$ & 3.477 $\pm$ 0.963\\ 
    $\theta/\pi$ & 0.872 $\pm$  0.051\\
\hline
\hline
\end{tabular}
\end{center}
\caption{Parameters of the model, obtained in Ref. \cite{avalos}.}
\label{parameters1}
\end{table}

Let us now explore the differential cross section as a function of the angular emission of one of the pions with respect to the collision axis, as a way to scan the relative energy between the $\omega$ and the $\rho$ resonances.
In order to calculate the differential cross section, we follow the kinematics as given in Ref. \cite{kumar} (A factor of $(2\pi)^9$ is added to agree with the phase space convention used by The Particle Data Group \cite{pdg}), which involves five Lorentz invariant variables:
    $s$ and $s_1$ defined above in addition to $ t_0 = (v_1 - p_1)^2$, $ u_1 = (q - p_2)^2 $ and $ t_1 = (v_1 - p_2)^2$.
The differential cross section at a given angle between the initial state particle $e^+ (v_1)$ and the final state particle $\pi^0(p_1)$ momenta (as seen from the center of mass frame) is given by


\begin{equation}
\begin{aligned}
    \frac{d\sigma(e^+e^-\to 2\pi^0\gamma)}{d\zeta} = & \frac{1}{512 \pi^4|\boldsymbol{v}_1| |\boldsymbol{v}_2| \lambda(s,m_e^2,m_e^2)^{1/2}} \int^{s_{1+}}_{s_{1-}} \frac{ds_1}{ (1-\xi_{1}^2)^{1/2}} \, \\
    &    \int_{u_{1-}}^{u_{1+}} \frac{du_1} {\lambda(s,m_\pi^2,u_1)^{1/2}(1-\eta_1^2)^{1/2}} \int_{t_{1-}}^{t_{1+}}\frac{dt_1}{(1-\zeta_1^2)^{1/2}} \, \overline{|\mathcal{M}|}^2,
\end{aligned}
\end{equation}

\noindent where $\zeta \equiv \cos \theta = \boldsymbol{p}_1 \cdot \boldsymbol{v}_1 / |\boldsymbol{p}_1| |\boldsymbol{v}_1|$ is the angle between the positron and the neutral pion momenta, which in the center of mass frame can be seen as the pion emission angle with respect to the collision axis, where the following kinematical definitions are considered
\begin{equation}\label{eq:6.11}
\begin{aligned}
    &\xi_1 = [s(s+m_\pi^2-s_1)-2s(m_e^2+m_\pi^2-t_0)][\lambda(s, m_e^2,m_e^2)\lambda(s,s_1,m_\pi^2)]^{-\frac{1}{2}},\\
    &\eta_1 = [2s(s_1+m_\pi^2)-(s + m_\pi^2-u_1)(s+s_1-m_\pi^2)][\lambda(s,m_\pi^2,u_1)\lambda(s,s_1,m_\pi^2)]^{-\frac{1}{2}},\\
    &\zeta_1 = (\omega_1 - \xi_1 \eta_1)[(1-\xi_1^2)(1-\eta_1^2)]^{-\frac{1}{2}},\\
    &\omega_1 = [s(s+m_\pi^2-u_1)-2s(m_e^2+m_\pi^2-t_1)][\lambda(s,m_e^2,m_e^2)\lambda(s,m_\pi^2,u_1)]^{-\frac{1}{2}}.
\end{aligned}    
\end{equation}
The limits of integration are
\begin{equation}\label{eq:6.10}
\begin{aligned}
    & s_{1-} = m_\pi^2, \quad s_{1+} = (\sqrt{s} - m_\pi)^2,\\
    & u_{1 \pm} = s + m_\pi^2 - \frac{(s_1 + m_\pi^2)(s+s_1-m_\pi^2)}{2s_1} \pm \frac{[\lambda(s,m_\pi^2,s_2)\lambda(s,s_1,m_\pi^2)]^{\frac{1}{2}}}{2s},\\
    & t_{1 \pm} = m_e^2 + m_\pi^2 - \frac{s + m_\pi^2 - u_1}{2} \pm \frac{[\lambda(s,m_e^2,m_e^2)\lambda(s,m_\pi^2,u_1)]^{\frac{1}{2}}}{2s} X_{1\pm},\\
    & X_{1\pm} = \xi_1 \eta_1 \pm \left[ (1-\xi_1^2)(1-\eta_1^2)\right]^{\frac{1}{2}}.
\end{aligned}    
\end{equation}

In order to fix the angle, the $t_0$ variable is turned into the $\zeta$ variable by
\begin{equation}
   t_0 = m_\pi^2 -2 (E_{v_1}E_{p_1} - \zeta |\boldsymbol{v_1}| |\boldsymbol{p_1}|),
\end{equation}
where    $ E_{v_1} = \frac{1}{2}\sqrt{s}$,
    $ E_{p_1} = \frac{m_\pi^2+s-s_1}{2\sqrt{s}}
    $, $ |\boldsymbol{v_1}| =|\boldsymbol{v_2}| = \frac{1}{2}\sqrt{s}
    $ and $ |\boldsymbol{p_1}| = \sqrt{E_{v_1}^2-m_\pi^2}$
 are obtained at the center of mass frame. This gives us the freedom to choose a specific value for $\zeta$ between $(-1,1)$, while fulfilling the limits of integration for $t_0$.

\begin{figure}
\centering
\includegraphics[scale=0.35]{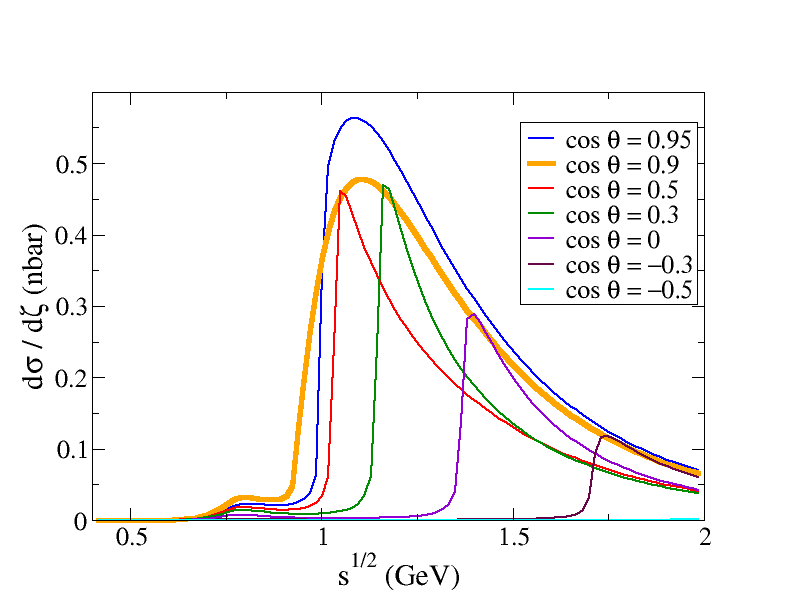}
\caption{\centering Differential cross section of the $e^+\,e^- \rightarrow \omega\pi^0 \rightarrow \pi^0\,\pi^0\,\gamma$ process for a set of values of $\zeta = \cos \theta$.}
\label{fig:dcross}
\end{figure}

\begin{figure}
\centering
\includegraphics[scale=0.40]{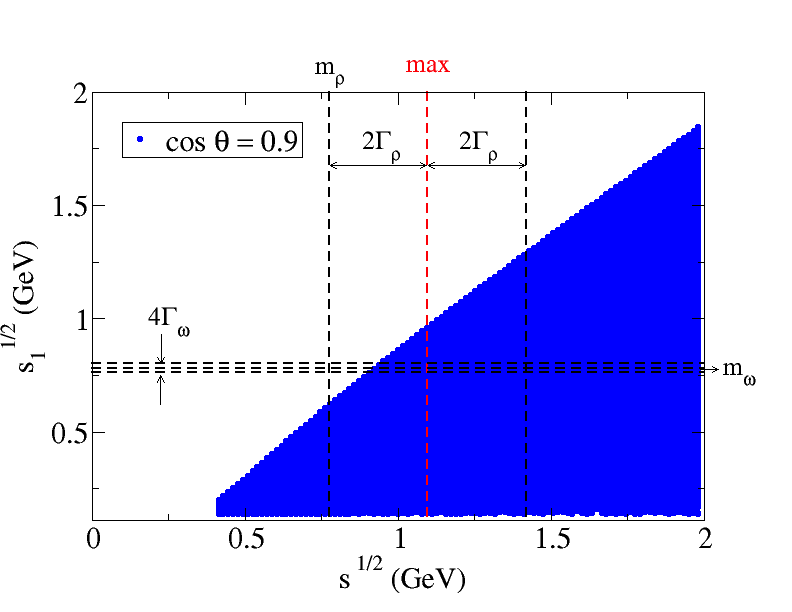}
\caption{\centering Dalitz region for $\sqrt{s}$ and $\sqrt{s_1}$ at $\zeta = 0.9$. The vertical and horizontal lines intersection defines the region where the $\omega$ and $\rho$ resonances have maximum interference, in units of their corresponding decay width.}
\label{fig:dalitz}
\end{figure}

For the sake of clarity let us consider, at this stage, only the $\rho$ contribution in the amplitude Eq. \ref{2pigamma}. The effect due to the $\rho^\prime$ will be included at the end. 
In Fig. \ref{fig:dcross} we show the differential cross section as a function of the center of mass energy $\sqrt{s}$ for a set of values of $\zeta$. We observe that it increases as it gets closer to $\zeta = 1$.
Negative values are highly suppressed. This is explained by picturing a final-state pion recoiling from the incident lepton trajectory. We consider the case for $\zeta = 0.9$ as a definite example to analyse.
In Fig. \ref{fig:dcross}, we plot this particular case (bold line), noticing the presence of two bumps; the first and small one at $\sqrt{s} \approx 0.78\,\, \text{GeV}$ and the second and big one at $\sqrt{s} \approx 1.097\,\, \text{GeV}$. The former coincides with the energy for the $\rho$ meson on-shell, $s= m_\rho^2$, but not the $\omega$. The latter corresponds to the $\omega$ meson on-shell which, since it is not explicitly dependent on $s$ but $s_1$, it is reflected at a higher energy.
A remaining question is, at which extent these two resonant contributions interfere with each other? For that purpose we explore the Dalitz region for $\sqrt{s}$ and $\sqrt{s_1}$ at $\zeta = 0.9$, as shown in Fig. \ref{fig:dalitz}. Analyzing this distribution we identify that the cross section resonates at $\sqrt{s_1} = m_\omega = 0.78266 \, \, \text{GeV}$ as it must be. At $\sqrt{s} = m_\rho$, the $\sqrt{s_1} = m_\omega$ condition is out of the region. However, at  $\sqrt{s} = 1.097 \,\, \text{GeV} \equiv max$  (maximum value identified from Fig. \ref{fig:dcross}) that condition can be reached. This explains the biggest bump, where both $\rho$ and $\omega$ particles resonant features combine to give a maximal enhancement. Although the $\rho$ is off-shell, its large decay width allows it to make a sizeable contribution. An indicator of the phase space effect is that the biggest bump starts rising at $\sqrt{s} \approx 0.93$ which intersects with $\sqrt{s_1} = m_\omega \,\, \text{GeV}$, corresponding with the opening of the $\omega - \pi$ states on-shell, but the maximum is reached at a higher energy. Measuring the energies in units of the corresponding decay width $\Gamma_\rho$ for $\sqrt{s}$ and $\Gamma_\omega$ for $\sqrt{s_1}$, we can identify that the maximum is 2$\Gamma$  away from the mass value. That is, it defines a rectangular region where both $\rho$ and $\omega$ resonant effects produce the maximum enhancement, as observed in Fig. \ref{fig:dcross}.

\begin{figure}
\centering
\includegraphics[scale=0.75]{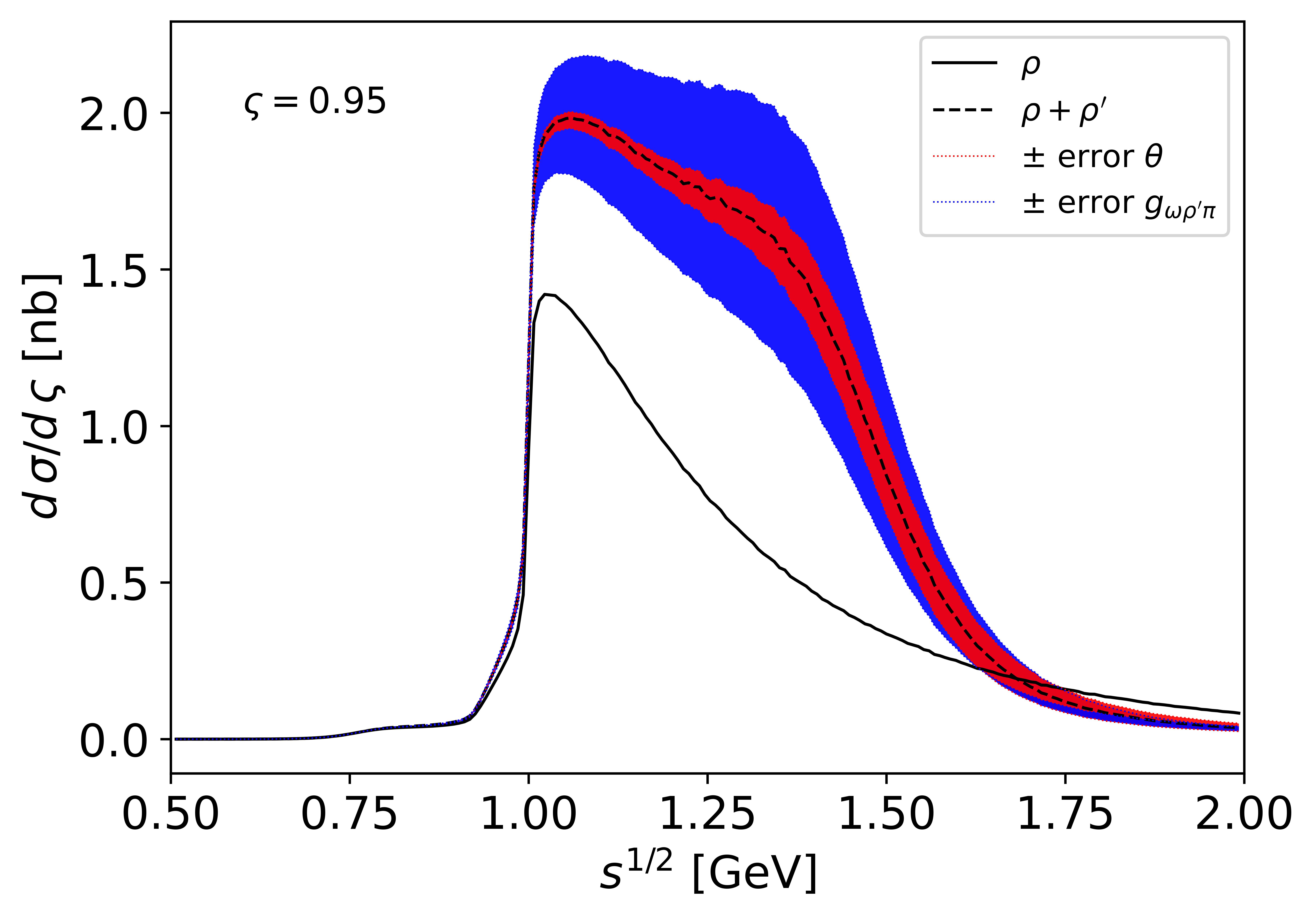}
\caption{\centering $e^+\,e^- \rightarrow \omega\pi^0 \rightarrow \pi^0\,\pi^0\,\gamma$ differential cross section at $\zeta = 0.95$. Considering the $\rho$ alone (solid line), and then adding the $\rho^\prime$ (dashed line). The broad band region corresponds to the uncertainty from $g_{\omega\rho^\prime\pi}$ and the narrow band to the $\theta$ phase uncertainty.}
\label{fig:dcrossall}
\end{figure}

A description of the differential cross section at $\zeta=0.95$ including the $\rho^\prime$ and the  region defined by the parameters uncertainty is shown in Fig. \ref{fig:dcrossall}. It is significantly sensitive to the $g_{\omega\rho^\prime\pi}$ coupling, the broad shaded region is defined by its uncertainty. The narrow shaded region corresponds to the relative phase parameter uncertainty. This suggest that by measuring the angular distribution in the region around 1.2 GeV, it may be possible to determine the resonant parameters involved with a better precision, namely the $g_{\omega\rho^\prime\pi}$ coupling constant and relative phase, $\theta$.
Notice that $m_{\rho^\prime}=1450$ MeV and $\Gamma_{\rho^\prime}=400$ MeV, makes the $\omega$ relatively closer to the $\rho^\prime$ than to $\rho$ in the context described above, using the decay width as a representative magnitude. Using a constant or an energy dependent width (modeled in a similar way to the $\rho$) makes no significant difference.

\begin{figure}
\centering
\includegraphics[scale=0.7]{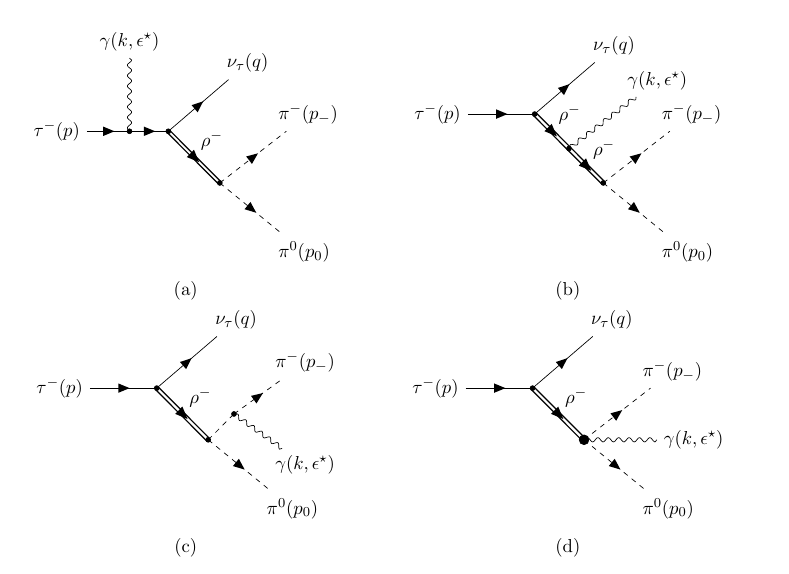}
\caption{\centering Feynman diagrams of the $\tau^{-}\to\pi^{-}\pi^{0}\nu_{\tau}\gamma$ decay corresponding to the structure independent part.}
\label{fig:taupipi}
\end{figure}

\section{Resonant enhancement in the $ \tau^{-}\rightarrow \pi^{-}\pi^{0}\nu_{\tau}\gamma$ decay}

In this section we describe the $\tau^- \to \pi^-\pi^0 \nu_\tau\gamma$ decay along the same lines as in \cite{Flores-Tlalpa:2005msx,FloresBaez:2006gf,FloresTlalpa:2006gs}. 
We explore the so-called $\omega$ channel dimeson invariant mass and angular distribution, similar in spirit to $e^+e^- \to \pi^0\pi^0 \gamma$ previously discussed. 
Then, we obtain the radiative correction function to the $\tau^- \to \pi^-\pi^0 \nu_\tau$ decay, $G_{EM}(t)$. Then, we compute the isospin symmetry breaking correction from this source to $\Delta a_\mu^{(HVP,LO)}$  paying attention to the uncertainties of the parameters involved.
 
Let us set the notation for the process as $ \tau^{-}(p) \rightarrow \pi^{-} \ (p_-) \ \pi^{0} \ (p_0) \ \nu_{\tau} (q)\gamma \ (k, \epsilon^*)$, where in parenthesis are the corresponding momenta and $\epsilon^*$ is the polarization vector of the photon. We define the auxiliary variables  $Q \equiv p_{0}-p_{-}$, $k_-\equiv p_{-}+p_{0} $ and $k_{+} \equiv k_-+k$ and the invariant variables $t=k^2_-$ and $t^{\prime}=k^2_+=t+2k_-\cdot k$.

The total amplitude for the $ \tau^{-}\rightarrow \pi^{-}\pi^{0}\nu_{\tau}\gamma$ process can be written in general as \cite{CEN,Bijnens:1992en}:
\begin{eqnarray}
{\cal{M}}_{T}&=&eG_FV^*_{ud}\epsilon^{*\mu}\Big[F_{\nu}\bar{u}(q)\gamma^{\nu}(1-\gamma_5)(m_{\tau}+\slashed p-\slashed k)\gamma_{\mu}u(p)\nonumber\\
&&\hspace{2 cm}+(V_{\mu\nu}-A_{\mu\nu})\bar{u}(q)\gamma^{\nu}(1-\gamma_5)u(p)\Big],\label{TCEN}
\end{eqnarray}
where the first line corresponds to the $\tau$ radiation, $F_{\nu} \equiv Q_\nu\frac{f_+[t]}{2\,p\cdot k}$ and $f_+[t]$ is the hadronic form factor obtained from the corresponding non radiative decay. $G_F$ is the Fermi constant and $V^*_{ud}$ the CKM matrix element. The $V_{\mu\nu}$ and $A_{\mu\nu}$ tensors correspond to the Vector and Axial contributions from the $W^-\rightarrow \pi^-\pi^0\gamma$ transition respectively (here $A_{\mu\nu}=0$ in accordance with previous analysis). $V_{\mu\nu}$ has the following structure:
\begin{eqnarray}
 V_{\mu\nu} &=&-f_+[t^{\prime}]\frac{p_{-\mu}}{p_-\cdot k}(Q-k)_{\nu}-f_+[t^{\prime}]g_{\mu\nu}\nonumber\\
&&+\frac{f_+[t^{\prime}]-f_+[t]}{k\cdot k_-}k_{-\mu}Q_{\nu} + \hat{V}_{\mu\nu},
\end{eqnarray}
where
\begin{eqnarray}
  \hat{V}_{\mu\nu} &\equiv&v_1\,p_-\cdot k\,F_{\mu\nu}(p_-) + v_2\, p_0\cdot k\, F_{\mu\nu}(p_0)\nonumber\\
 &&+v_3\,p_0\cdot k\,p_-\cdot k\,L_{\mu}(p_-,p_0)\,p_{-\nu} + v_4\,p_0\cdot k\,p_-\cdot k L_{\mu}(p_-,p_0)k_{+\nu},\label{Vmunu}
\end{eqnarray}
and we have made use of the following functions:
\begin{eqnarray}
L_{\mu}(a,b) &\equiv& \frac{a_{\mu}}{a\cdot k}-\frac{b_{\mu}}{b\cdot k},\nonumber\\
F_{\mu\nu}(a) &\equiv& g_{\mu\nu} -\frac{a_{\mu}k_{\nu}}{a\cdot k}.\nonumber
\end{eqnarray}
The $v_i$ functions are determined from the specific model considered for the hadronic description. In our case, given by the Lagrangian Eq. (\ref{Lvmd}), in addition to the vector meson - photon interaction ($VV\gamma$), which is taken in analogous way as for the W gauge boson ($WW\gamma$) incorporating the finite width effect in a gauge invariant way \cite{LopezCastro:1999dp}. The structure independent (SI) diagrams are depicted in Fig. \ref{fig:taupipi}, which include MI and MD parts, this last associated to the $\rho$ meson MDM ($\beta_0$) and taken to be $\beta_0=2$ in $e/2m_\rho$ units. The weak $\rho$ coupling is set to $G_\rho=\sqrt{2}m_\rho^2/g_{\rho\pi\pi}$. The $v_i$ functions are given by:
\begin{eqnarray}
 v_1 &=& -v_2 = \beta_0\frac{[f_+(t^{\prime})-f_+(t)]}{2\,k_-\cdot k},\nonumber\\
  v_3 &=&0,\\
 v_4 &=&2\,\Big(\frac{\beta_0}{2}-1\Big)\frac{(1+i\Gamma_\rho/m_\rho)}{m^2_{\rho}}\frac{[f_+(t^{\prime})-f_+(t)]}{k_-\cdot k},\nonumber
 \label{vis}
\end{eqnarray}
where we have used
$ \frac{(f_+[t^{\prime}]-f_+[t])}{2\,k_-\cdot k} = (1+i\Gamma_\rho/m_\rho)\,\frac{f_+[t]\,f_+[t^{\prime}]}{m^2_{\rho}}$.
We identify the MI part, in accordance to the Low theorem, as those contributions of order  $O(k^{-1})$ and $O(k^{0})$ \cite{Low}:
\begin{eqnarray}
 {\cal{M}}_{Low}&=& eG_FV^*_{ud}\epsilon^{*\mu}\,\Big\{f_+[t]\,L_\mu(p,p_-)Q_{\nu}+ 2\,p_0\cdot k \,L_\mu(p_0,p_-)\,\frac{df_+[t]}{dt}\,Q_{\nu}\nonumber\\
&&-\frac{f_+[t]}{2\, p\cdot k}\,\Big[F_{\mu\nu}(Q)Q\cdot k + iQ^{\alpha}k^{\beta}\epsilon_{\nu\alpha\beta\mu}\Big]\nonumber\\
&&-f_+[t]\,F_{\mu\nu}(p_-)\Big\}\,l^{\nu},\label{MLow}
\end{eqnarray}
where $l^{\nu}= \bar{u}(q)\gamma^{\nu}(1-\gamma_5)u(p)$.
This is the same  result obtained previously in the VMD \cite{FloresBaez:2006gf,FloresTlalpa:2006gs} and chiral perturbation theory ($\chi_{PT}$) \cite{CEN,pablo20} descriptions, with $\hat{V}_{\mu\nu}$ and $A_{\mu\nu}$ null.

The form factor $f_+[t]$ is obtained from a fit to the two pion invariant mass distribution of the non radiative decay, measured by Belle \cite{Belle:2008xpe}. It includes the $\rho(770)$, $\rho(1450)$ and $\rho(1700)$ vector mesons by:
\begin{equation}
    f_+[t] = \frac{1}{1+\beta+\gamma}\,\Big\{f_{\rho}[t] + \beta \,f_{\rho^{\prime}}[t] + \gamma\,f_{\rho^{\prime\prime}}[t] \Big\},\label{GS}
\end{equation}
where $\beta = B_0\, e^{i\,f_b} $ and $ \gamma  = G_0\, e^{i\,f_g}$.
The parameters are listed in Table \ref{tab:fitff}
\begin{table}[h!]
    \centering
    \begin{tabular}{cccc}
        Parameter & Value & Parameter & Value \\ \hline \hline
        $m_{\rho}$ & 0.7747 GeV & $\Gamma_{\rho}$ & 0.14612 GeV \\
        $m_{\rho^{\prime}}$ & 1.3832 GeV & $\Gamma_{\rho^{\prime}}$ & 0.5653 GeV\\
        $m_{\rho^{\prime\prime}}$ & 1.868 GeV & $\Gamma_{\rho^{\prime\prime}}$ & 0.3941 GeV\\
        $B_0$ & -0.4028 & $f_b$ & 1.1321 \\
        $G_0$ & -0.1725 & $f_g$ & $4.3756 \times 10^{-8}$ \\
        \hline
    \end{tabular}
    \caption{Parameters obtained from a fit to the Belle data form factor $f_+[t]$.}
    \label{tab:fitff}
\end{table}

The involved couplings from the model are related to the fit by:
\begin{equation}
    \frac{\beta}{1+\beta+\gamma} = \frac{m^{2}_{\rho}}{m^{2}_{\rho^{\prime}}}\,\frac{G_{\rho^{\prime}}\,g_{\rho^{\prime}\pi\pi}}{G_{\rho}\,g_{\rho\pi\pi}},
    \hspace{0.5 cm}\frac{\gamma}{1+\beta+\gamma} = \frac{m^{2}_{\rho}}{m^{2}_{\rho^{\prime\prime}}}\,\frac{G_{\rho^{\prime\prime}}\,g_{\rho^{\prime\prime}\pi\pi}}{G_{\rho}\,g_{\rho\pi\pi}},\label{bg}
\end{equation}
where $G_{\rho}$, $G_{\rho^{\prime}}$ and $G_{\rho^{\prime\prime}}$ are the corresponding vector mesons weak couplings. Notice that only the ratios are involved and fixed by the fit parameters. A comparison of the form factor with respect to the dispersion relation result can be seen in \cite{LopezCastro:2015cja}.

\begin{figure}
\centering
\includegraphics[scale=0.5]{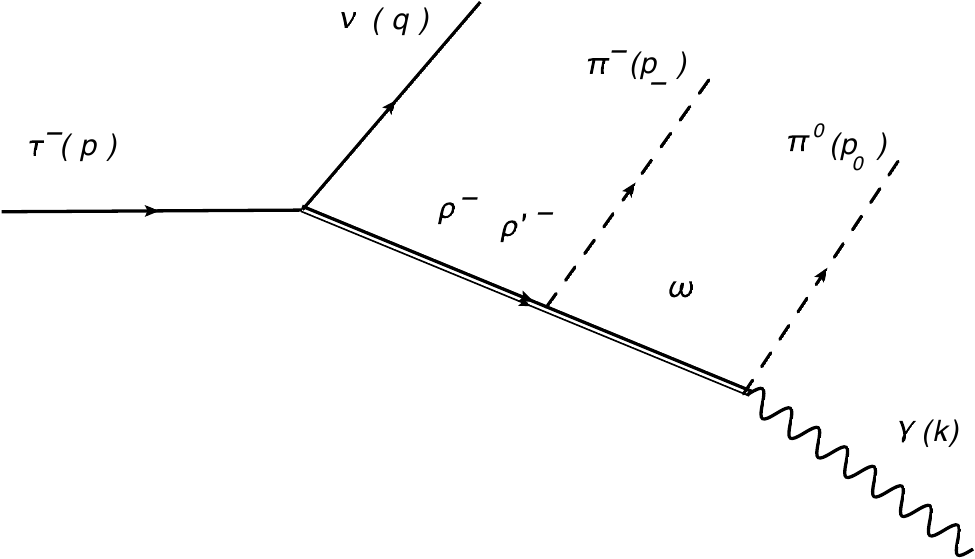}
\caption{\centering Contribution to the $\tau^- \to \pi^-\pi^0 \nu_\tau \gamma$ decay, driven by the $\omega$.}
\label{Fig:omega}
\end{figure}

Now, we proceed to analyze the MD part coming from the $\omega$ channel, depicted in Fig. \ref{Fig:omega}. There, we show the diagram for the $\tau^- \to \pi^-\pi^0 \nu_\tau \gamma$ decay, driven by the presence of the $\rho$ and $\omega$ intermediate states. It has been shown that this is the only MD relevant channel \cite{FloresBaez:2006gf,FloresTlalpa:2006gs}. The amplitude can be written as:
\be
{\cal M}_\omega=eG_F V^*_{ud}\frac{G_\rho}{\sqrt{2}} \frac{g_{\omega\rho\pi}^2 e^{i\theta_w}}{g_\rho m_\rho^2 m_\omega^2} f_\omega[r]
f_o[t^\prime]
{\epsilon_{\alpha\sigma\mu}}^\lambda
{\epsilon_{\phi\lambda\chi}}^\nu
k^\sigma p_0^\alpha (p_0+k)^\phi p_-^\chi \epsilon^{*\mu}  \ell^{\nu},
 \label{omega}
\ee
where $r\equiv (p_0+k)^2$ and $f_o[t^\prime]$ includes the $\rho$ and $\rho^\prime$ contributions
\be
f_o[t^\prime]\equiv
\frac{1}{1+B_{1} e^{i\theta}}\,\left\{
f_\rho[t^\prime]+B_1 e^{i\theta}  f_{\rho^\prime}[t^\prime] \right\}.
\ee
The parameter $B_1$ is related to the coupling constants of the model by $B_1= |(m_\rho/m_\rho^\prime)^2 (G_{\rho^\prime}/G_\rho)(g_{\omega\rho^\prime\pi}/g_{\omega\rho\pi})|$ (with $G_{\rho^\prime}/G_\rho$ determined from the parameters of $f_+[t]$) and $\theta$ is the relative phase between the $\rho$ and $\rho^\prime$ contribution to the $\omega$ channel. This strong phase has the same origin as in $e^+e^- \to \pi^0 \pi^0 \gamma$, and therefore is assumed to be the same. Global phase effects may be different compared to the $e^+e^-$ mechanism. The relative phase between the channel itself, encoded in $\theta_w$, and the SI amplitude is taken to be positive.  Thus $f_o[t^\prime]$, although similar in structure to $f_+[t]$ (without the $\rho^{\prime\prime}$) involves different values for the parameters associated to the $\rho^\prime$ contribution, determined in a previous analysis \cite{avalos}.

The amplitude can be set in the general structure form, Eq. (\ref{TCEN}), as:
\be
{\cal M}_\omega=eG_F V^*_{ud}
\epsilon^{*\mu} \hat{V}^{(\omega)}_{\mu\nu}  \ell^{\nu},
 \label{omegav}
\ee
and $\hat{V}^{(\omega)}_{\mu\nu}$ contributes to $\hat{V}_{\mu\nu}$ with the following coefficients:
\begin{eqnarray}
 v_1^{\omega} &=& -C_\omega f_\omega[r] f_o[t^\prime] (p_0+2k)\cdot p_0, \nonumber\\
v_2^{\omega}&=&C_\omega f_\omega[r] f_o[t^\prime] (p_0+k)\cdot p_-,\\
 v_3^{\omega} &=&C_\omega f_\omega[r] f_o[t^\prime] ,\nonumber\\
 v_4^{\omega} &=&-C_\omega f_\omega[r] f_o[t^\prime], \nonumber
\end{eqnarray}
where 
$C_\omega= g_{\omega\rho\pi}^2/( m_\omega^2 g_\rho  g_{\rho\pi\pi})$.

In order to evaluate the corresponding contributions we use the values for the couplings obtained from the parameter analysis \cite{avalos}, Table \ref{parameters1}.

\subsection{Pion angular distribution}

\begin{figure}
\centering
\includegraphics[scale=0.75]{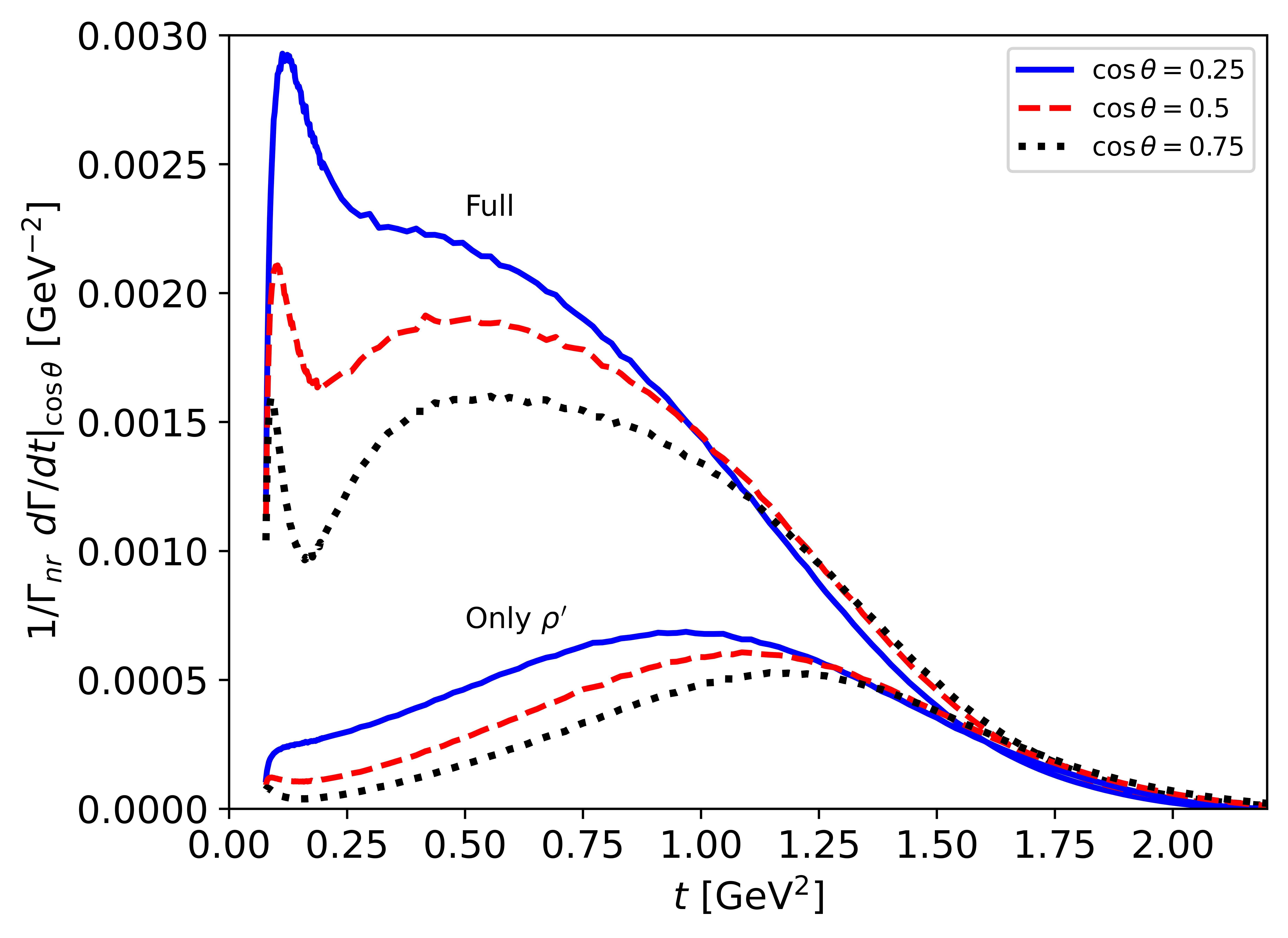}
\caption{Dimeson invariant mass distribution due to the $\omega$ channel, normalized to the non-radiative decay width ($\Gamma_{nr}$) for several angles of the charged pion emission. Lines in the upper region of the figure (Full) consider $\rho$ and $\rho^\prime$. The lines in the lower region consider only the $\rho^\prime$ contribution, for the corresponding angles.}
\label{dimesonangle}
\end{figure}

The dipion invariant mass distribution has been shown to be a useful observable to study the underlying dynamics  of $\tau^- \to \pi^-\pi^0 \nu_\tau \gamma$ decay \cite{CEN,Flores-Tlalpa:2005msx,pablo20}. The distribution associated to a particular angular emission of the charged pion with respect to the dipion momenta in the $\tau$ rest frame may resemble the behavior observed in the $e^+e^- \to \pi^0 \pi^0 \gamma$ process discussed previously.
In Fig. \ref{dimesonangle}, we show the dimeson invariant mass distribution due to the $\omega$ channel, normalized to the non-radiative decay width ($\Gamma_{nr}$) for several angles of the charged pion emission, obtained using the same kinematics as in Ref. \cite{Flores-Tlalpa:2005msx}. Lines in the upper region of the figure (Full) consider $\rho$ and $\rho^\prime$. The lines in the lower region consider only the $\rho^\prime$ contribution, for the corresponding angles. We observe that small angles are favored and the individual resonant structures are split. 

\begin{figure}
\centering
\includegraphics[scale=0.75]{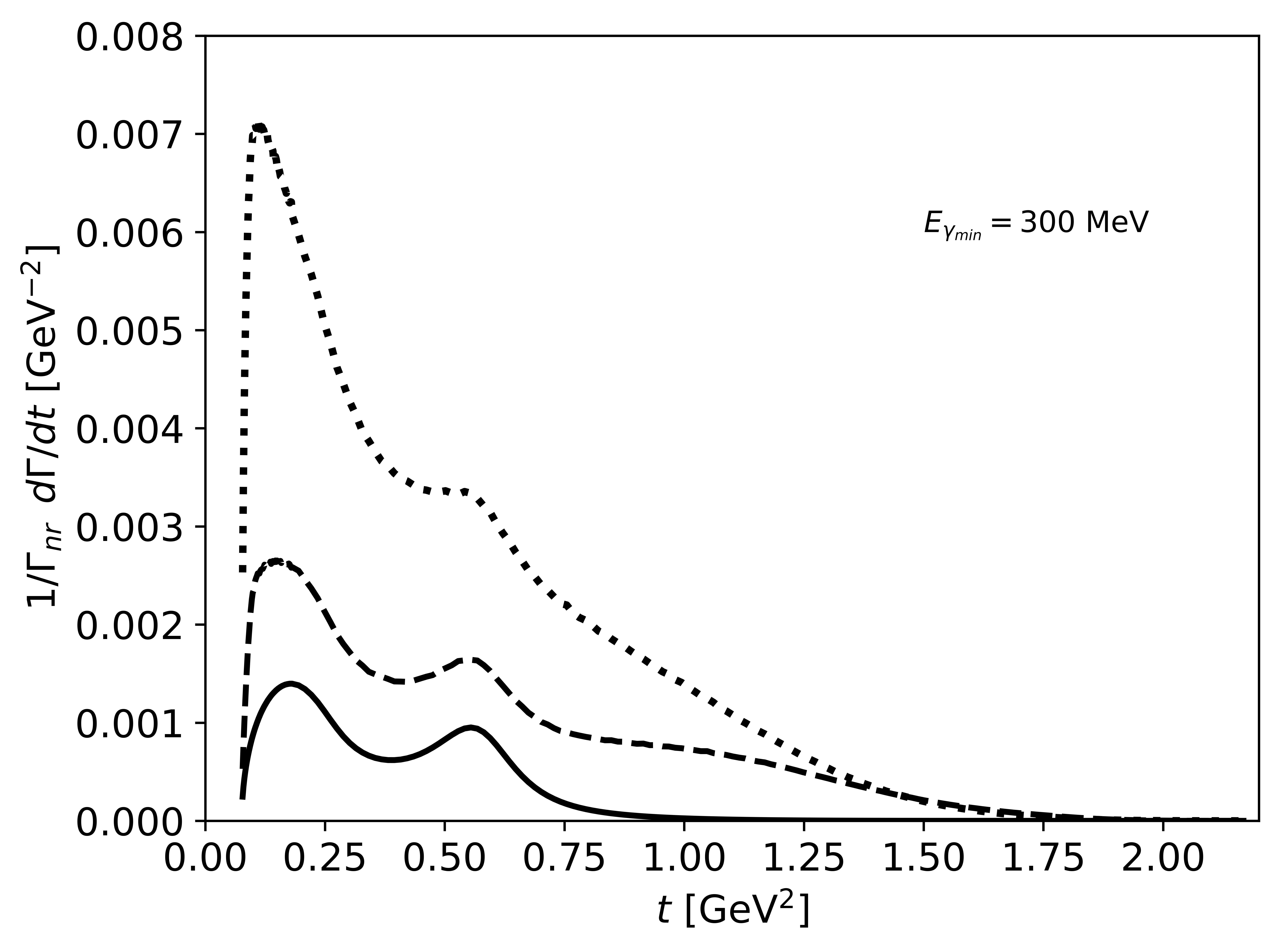}
\caption{Dipion invariant mass distribution using a cut off of ${E_{\gamma}}_{min}=300$ MeV. The dotted line corresponds to the total dipion invariant mass, the dot-dashed line is the contribution excluding the $\rho^\prime$ in the omega channel, and the solid line is the SI contribution.}
\label{dimesonfull}
\end{figure}

In Fig. \ref{dimesonfull}, we show the dipion invariant mass distribution regardless of the angle. The dotted line corresponds to the total dipion invariant mass (obtained from the SI diagrams plus interference with the $\omega$ channel), the dot-dashed line is the contribution excluding the $\rho^\prime$ in the $\omega$ channel, and the solid line is the SI contribution.
We use a cut off for the photon energy of ${E_{\gamma}}_{min}=$300 MeV, implemented by introducing a fictitious mass at the kinematical level, such that the photon energy can not go lower than that energy.

\subsection{Correction to the $\tau$-based muon MDM estimate}

The muon MDM estimate, based on $\tau$ data, requires to incorporate the correction from all the contributions that break the conserved vector current (CVC) hypothesis. In particular, to determine the leading hadronic contribution, $a_\mu^{(HVP,LO)}$, from the two pions decay mode, requires to incorporate the correction, $\Delta a_\mu^{(HVP,LO)}$, from all the the isospin symmetry breaking sources, denoted by $R_{IB}(t)$, with $t$ the  dipion invariant mass square:
\begin{equation}
    R_{\mathrm{IB}}(t) = \frac{FSR\,(t)}{G_{EM}(t)}\,\frac{\beta^{3}_{\pi^{+}\pi^{-}}}{\beta^{3}_{\pi^{+}\pi^{0}}}\,\Big|\frac{F_V[t]}{f_+[t]}\Big|^{2},
\end{equation}
where $FSR\,(t)$ accounts for the final state radiation from the pions, $G_{EM}(t)$ is the electromagnetic radiative correction function, $\beta^{3}_{\pi^+\pi^-}/\beta^{3}_{\pi^{+}\pi^{0}}$ is the phase space factor correction and $|F_V[t]/f_+[t]|^{2}$ is the form factor correction from the charged ($f_+[t])$) with respect to the neutral ($F_V [t]$) one. These corrections have been computed, with the main source of uncertainty coming from the form factors ratio and the electromagnetic term \cite{Aoyama:2020ynm,CEN,Davier:2010nc,pablo20,Davier:2010fmf,Jegerlehner:2017gek,
Benayoun:2012etq,Benayoun:2021ody}.

Here, we focus on the correction to $a_\mu^{(HVP,LO)}$ from $G_{EM}(t)$, which is estimated by \cite{CENPLB,CEN}:
\begin{eqnarray}
\Delta a_\mu^{(HVP,LO)}\vert_{G_{EM}(t)}&=&\frac{1}{4\pi^3}\int^{t_{max}=m_\tau^2}_{t_{min}=4m^2_{\pi}} dt \, K(t)\frac{K_{\sigma}(t)}{K_{\Gamma}(t)}\frac{d\Gamma_{2\pi(\gamma)}}{dt}\nonumber\\
&&\times \Big[\frac{1}{G_{EM}(t)}-1\Big],
\end{eqnarray}
where $K(t)$ is the QED Kernel function, given by
\begin{equation}
    K(t) = \frac{x^{2}}{2}(2-x^{2})+\frac{(1+x^{2})(1+x)^{2}}{x^{2}}\,\Big(\mathrm{ln}(1+x)-x+\frac{x^{2}}{2}\Big)+\frac{(1+x)}{(1-x)}x^{2}\mathrm{ln}(x),
\end{equation}
where
\begin{equation}
x=\frac{1-\beta_{\mu}}{1+\beta_{\mu}},\hspace{1 cm} \beta_{\mu} = \sqrt{1-4m^{2}_{\mu}/t},
\end{equation}
\begin{equation}
K_{\Gamma}(t) = \frac{G^{2}_F\,|V_{ud}|^{2}\,m^{3}_{\tau}}{384\,\pi^{3}}\,\Big(1-\frac{t}{m^{2}_{\tau}}\Big)^{2}\,\Big(1+\frac{2\,t}{m^{2}_{\tau}}\Big),\qquad \mbox{and}\quad K_{\sigma} = \frac{\pi\,\alpha^{2}}{3\,t}.
\end{equation}

\begin{figure}
\includegraphics[scale=1,angle=0]{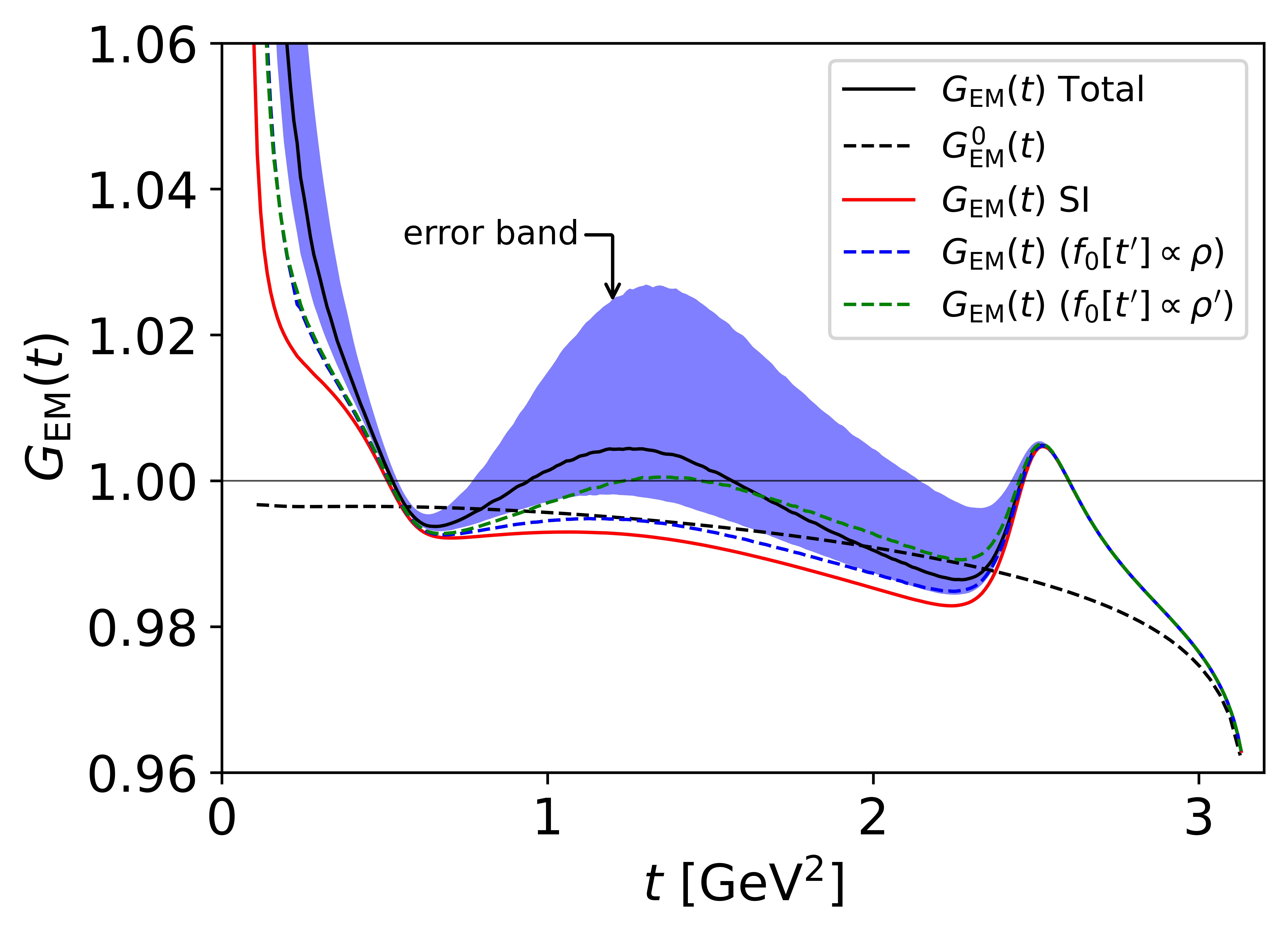}
\caption{$G_{EM}(t)$ function including several contributions: Total (black solid line)  is the SI contribution and the interference with the $\omega$ channel considering only the $\rho$. Adding the $\rho^\prime$ and using the current uncertainties on the parameters defines the shaded region. The contribution from only the $\rho$ in the $\omega$ channel (blue dashed line) and the contribution from only the $\rho^{\prime}$ in the $\omega$ channel (green dashed line), the SI contribution (red solid line) and the result for $G^0_{EM}(t)$ (black dashed line).}
\label{Gemto}
\end{figure}

This contribution, due to the lack of experimental information, is estimated theoretically by considering the virtual and real photon emission in the $\tau^- \to \pi^-\pi^0 \nu_\tau \gamma$ decay.
The $\omega$ contribution enters through the interference with the SI Bremsstrahlung \cite{Davier:2010fmf} and has been studied considering only the $\rho$ in the $\omega$ channel. Here, we extend the analysis to incorporate the $\rho^\prime$, which is already far from the soft photon approximation regime and requires to consider the results with caution as they are fully model dependent. Still, we do it in an attempt to explore the role of the parameters involved.\\
Let us recall the general procedure to compute the electromagnetic correction:
The photon inclusive dipion invariant mass distribution at $O(\alpha)$ can be set, in terms of the non-radiative decay, $\Gamma^0_{2\pi}$, as \cite{CEN}
\begin{equation}
\frac{d\Gamma_{2\pi(\gamma)}}{dt} = \frac{d\Gamma^0_{2\pi}}{dt}G_{EM}(t),
\end{equation}
where $G_{EM}(t)$ encodes the long distance radiative corrections.
In general, the electromagnetic function can be split into two parts \cite{CEN,FloresBaez:2006gf}:
\begin{equation}
G_{EM}(t) = G^0_{EM}(t) + G^{rest}_{EM}(t),
\label{Gemtsplit}
\end{equation}
where $G^0_{EM}(t)$ accounts for the virtual and real contribution up to $O(k^{-2})$, and $G^{rest}_{EM}(t)$ includes the remaining higher order contributions from the real part. $G^0_{EM}(t)$ has been computed in \cite{CEN} and $G^{rest}_{EM}(t)$, which includes MI and MD parts, has been computed  in two frameworks, $\chi_{PT}$ \cite{CEN,pablo20} and VMD \cite{Flores-Tlalpa:2005msx,FloresBaez:2006gf,FloresTlalpa:2006gs}, as mentioned before.
 In Fig. \ref{Gemto} we show the electromagnetic function including different contributions. Total (black solid line) corresponding to the SI and interference with the $\rho$ part of the $\omega$ channel. The uncertainties associated are not visible at the current scale, that is, at this stage the MD contribution is well settled. Adding the $\rho^\prime$ and using the current uncertainties on the parameters defines the shaded region, signaling the lack of precision on such contribution. We have also plotted the contribution only from the $\rho^\prime$ in the $\omega$ channel (green dashed line), the SI contribution (solid red line) and the result for $G^0_{EM}(t)$ (black dashed line).

\begin{figure}
\includegraphics[scale=1,angle=0]{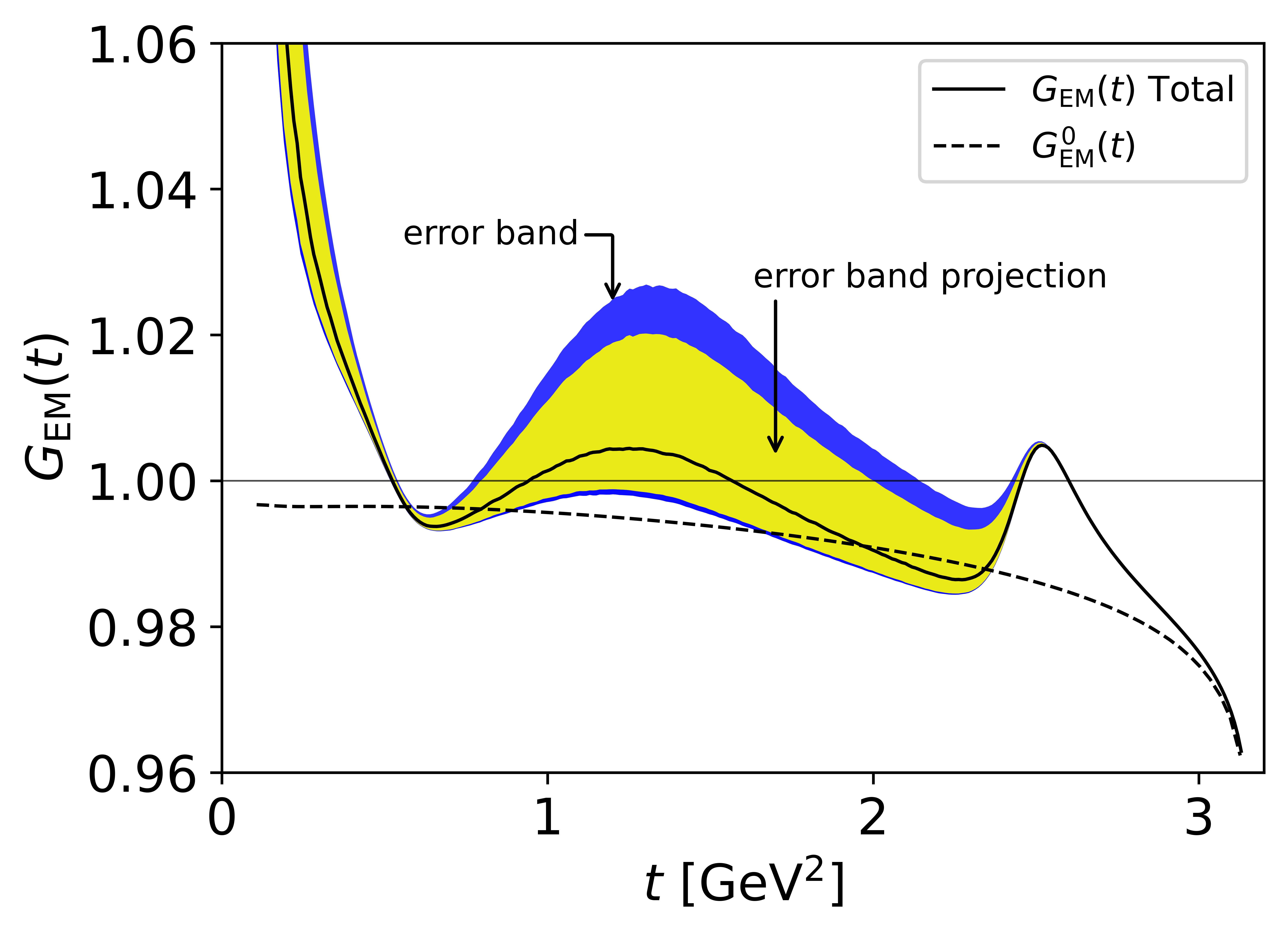}
\caption{$G_{EM}(t)$ function for the current uncertainties on the  $\rho^\prime$ parameters (broad shaded region) as in Fig. \ref{Gemto} and the projection region (inside region) considering an improvement on the $g_{\omega\rho^\prime\pi}$ of 20\%. We also include the result for $G^0_{EM}(t)$.}
\label{Gemtprojection}
\end{figure}

 In Fig. \ref{Gemtprojection}, we show the  electromagnetic function for the current uncertainties on the $\rho^\prime$ parameters (broad shaded region), as in Fig. \ref{Gemto}, and the projection region (inside region) considering an improvement on the $g_{\omega\rho^\prime\pi}$ of 20\%, which may be attainable by measuring the $e^+e^- \to \pi^0\pi^0\gamma$ angular distribution described in the first part of this work.

 Following the form of $G_{EM}(t)$ as in Eq. (\ref{Gemtsplit}), we can compute the contributions to $\Delta a_\mu^{(HVP,LO)}|_{G_{EM}(t)}$ from the different terms. Namely, $G_{EM}^0(t)$ and then adding $G_{EM}^{rest}(t)$  parts. The numerical integration is performed in the region from $t_{min} = 0.0773$ GeV$^2$ to  $t_{max} = 3.14$ GeV$^2$.
In Table \ref{amutabla}, we show the results considering the different contributions, namely:\\
$(i)$ $G_{EM}^0(t)$;\\
$(ii)$ $G_{EM}(t)$(SI), the SI part in addition to $G_{EM}^0(t)$;\\
$(iii)$ $G_{EM}(t)$ (Full), the SI plus the $\rho$ contribution in the $\omega$ channel in addition to $G_{EM}^0(t)$;\\
$(iv)$ $G_{EM}(t)$ (Full+$\rho^\prime $ ), similar to the previous case but adding the $\rho^\prime $ contribution in the $\omega$ channel.\\
$(v)$ $G_{EM}(t) $ (Projection) is the result for a projected reduction of 20\% in the $g_{\omega\rho^\prime\pi}$ uncertainties, while keeping the central value fixed.
The uncertainties are taken to account for the corresponding individual parameters uncertainties, assumed  uncorrelated. \\
The results here obtained for $\Delta a_\mu^{(HVP,LO)}\vert_{G_{EM}(t)}$ considering $(i)$ and $(ii)$ are consistent with the ones obtained in previous works, for example in \cite{CEN,CENPLB}. The result considering $(iii)$  is consistent with previous estimates  $-37 \times 10^{-11}$ \cite{FloresBaez:2006gf,FloresTlalpa:2006gs,Davier:2010fmf}. This large contribution from the $\omega$ channel is well under control with relatively small uncertainties mainly associated to the $g_{\omega\rho\pi}$ coupling.
The result considering $(iv)$ becomes anomalously large and may signal the break of the approach, and would call for further analysis, we have pointed out the origin of the main uncertainties to the $\rho^\prime$ and its interaction with the $\omega$ through the $g_{\omega\rho^\prime\pi}$ coupling. It is close to  $-(76 \pm 46) \times 10^{-11}$  obtained at $O(p^6)$ in a Chiral description with resonances \cite{pablo20}.\\
For comparison purposes, we can consider the total contribution to $\Delta a_\mu^{(HVP,LO)}$ from the rest of the isospin symmetry breaking terms in $R_{IB}(t)$ and the SD electroweak radiative correction $S_{EW}$, as obtained in \cite{Davier:2010fmf}. This would imply a shift in the total $\Delta a_\mu^{(HVP,LO)}$ from $-(16.07 \pm 1.85)  \times 10^{-11}$ to  $-(18.0 \pm 1.69) \times 10^{-10}$ considering only the $\rho$ in the $\omega$ channel, and  $-(23.55 ^{+3.6}_{- 9.34}) \times 10^{-10}$ when adding the $\rho^\prime$.

\begin{table}
\begin{tabular}[t]{|cc |c|}
\hline
 &$G_{EM}(t)$  & $\Delta a_\mu^{(HVP,LO)}$  ( $ \times 10^{-11}$)  \\
\hline
$(i)$ &  $G_{EM}^0(t)$ & 18.3\\
$(ii)$ & $G_{EM}(t)$(MI) & -12.03   \\
$(iii)$ & $G_{EM}(t)$(SI) & -14.8   \\
$(iv)$ & $G_{EM}(t)$ (Full) & $-38.51^{+4.04}_{-4.83}$ \\
$(v)$ & $G_{EM}(t)$ (Full+$\rho^\prime $) & $-94.03^{+32.2}_{-92.04}$ \\
$(vi)$ & $G_{EM}(t) $ (Projection) & $-94{.}03^{+28.15}_{-66.22}$\\
\hline
\end{tabular}
\caption{ $\Delta a_\mu^{(HVP,LO)}|_{G_{EM}(t)}$  ( $ \times 10^{-11}$) for several contributions of $G_{EM}(t)$.}
\label{amutabla}
\end{table}

\section{Conclusions }
We have explored the enhancement mechanism due to the resonant properties of the $\rho$ and $\omega$ mesons, when such resonances carry different momenta, to exhibit how both resonances combine to produce the enhancement. First, we considered the $e^+e^- \to \pi^0 \pi^0 \gamma$ process and made use of the differential cross section at a given angle of emission of one of the pions, to tune the individual features of the two resonances. There, we found that the main combined resonant contribution takes place when both are within the energy region defined by $m_V\pm 2\Gamma_V$. 
Then, we incorporated the $\rho^\prime$ and showed that, it becomes important early in the energy region, with respect to its mass, thanks to the same enhancement mechanism between the $\rho$ and the $\omega$. We identified the sensibility to two parameters of the $\rho^\prime$, namely the relative phase with respect to the $\rho$ and the $g_{\omega\rho^\prime\pi}$ coupling. The angular distribution proved to be a scenario where this last can be determined with improved precision.
In a second step, we considered the radiative $\tau^- \to \pi^-\pi^0 \nu_\tau\gamma$  decay, whose main MD contribution, the $\omega$ channel, exhibits similar features to  $e^+e^- \to \pi^0 \pi^0 \gamma$. Thus, following the same approach, we showed that the dipion invariant mass distribution at particular angles of the charged pion emission is sensitive to the individual resonant states. We computed the interference of this  channel with the known dominant SI contribution, and obtained the electromagnetic function $G_{EM}(t)$, this is found to be well settled in the soft photon approximation regime, dominated by the $\rho$ meson. A large source of uncertainty was identified upon the inclusion of the $\rho^\prime$, described in a similar way as in $e^+e^- \to \pi^0 \pi^0 \gamma$. 
We obtained the electromagnetic correction to the muon MDM estimate. The leading contribution is in accordance with previous determinations regardless of the model. The MD part involves two sources, the $\rho$ meson MDM whose value we fixed to $\beta_0=2$, and the so-called $\omega$ channel, being this last the main contribution. Our results confirm the previous finding \cite{FloresBaez:2006gf,FloresTlalpa:2006gs} that a large MD effect is at play and is the reason of the observed deviation with respect to the Chiral approach at $O(p^4)$ \cite{CEN}. The contribution of the $\omega$ channel have relatively small uncertainties considering only the $\rho$ meson and becomes anomalously large upon the inclusion of the $\rho^\prime$ (with also large uncertainties). In view of the soft photon approximation, this may point out to a possible breaking of the approach. Estimates using the $\chi_{PT}$ with resonances at $O(p^6)$ \cite{pablo20}, found out that the higher order terms were important, pointing out to the relevance of the $\omega$ and other contributions, although with a different handling of the uncertainties due to the model approach.\\
The form factor used to compute $G_{EM}(t)$ by definition appears in the numerator and denominator. Thus, its effect becomes subdominant and should not make difference in the results obtained above. Also isospin symmetry breaking associated to the neutral and charged pion mass difference, within the radiative process, is subdominant and thus its effect on $\Delta a_\mu^{(HVP,LO)}$ is negligible.
For the $\omega$ decay width we made use of a constant width, based on the fact that the main contribution is for energies around the $\omega$ mass. Corrections from an energy dependent width are expected for off-shell $\omega$, mainly from the opening of the $\omega \to \rho \pi$ channel. This is particular important for the precision estimate of g-2 contribution from $e^+e^- \to \pi^0 \pi^0 \gamma$. We have neglected this and other effects such as the $\phi$ meson, where the same consideration about the width takes place \cite{Moussallam:2013una,Moussallam:2021dpk,jorge}.
For the radiative $\tau$ decay correction, which is already subleading, this effect is expected to be also negligible in general grounds, we are not aware of any particular work on this aspect.

We would like to conclude stating that the link between the $e^+e^- \to \pi^0 \pi^0 \gamma$ process and the  $\omega$ channel of the $\tau^- \to \pi^-\pi^0 \nu_\tau\gamma$ decay, that is the double pole resonant enhancement, can be used to gain further insight into the description of such processes and that there are particular scenarios where we can profit from this effect.

\begin{acknowledgments}
We acknowledge the support of CONACyT, Mexico Grant No. 711019 (A. R.) and the support of DGAPA-PAPIIT UNAM, under Grant No. IN110622, PRIDIF IFUNAM fellowship (A. R.). We thank Doctor Gabriel L\'opez Castro and Doctor Pablo Roig for very useful discussions and comments.
\end{acknowledgments}

%

\end{document}